\documentstyle[preprint,aps]{revtex}
\tightenlines

\begin{document}

\draft
\title
{Hausdorff dimension and filling factor}

\author
{Wellington da Cruz\footnote{E-mail: wdacruz@fisica.uel.br}}

\address
{Departamento de F\'{\i}sica,\\
 Universidade Estadual de Londrina, Caixa Postal 6001,\\
Cep 86051-970 Londrina, PR, Brazil\\}
 
\date{\today}

\maketitle

\begin{abstract}
We propose a {\it new hierarchy scheme} for the 
{\it filling factor}, a 
parameter which characterizes the occurrence of the
 Fractional
 Quantum Hall 
Effect ( FQHE ). We consider the Hausdorff dimension, 
$h$, as a
 parameter for classifying 
 fractional spin particles, such 
that, it is
 written in terms of the statistics of the collective 
 excitations. The 
 number $h$ classifies these excitations with different 
 statistics 
 in terms of its homotopy class.

\end{abstract}



In a series of papers\cite{R1}, we have obtained a set 
of results 
with respect to fractional spin particles. Now we make 
the connection 
with the filling factor, a parameter that appears 
in the context 
of the FQHE and for it we have experimental 
values\cite{R2}. That 
connection can be made once the anyonic model has 
been considered to 
explain this phenomenon\cite{R3}.

The FQHE is associated with a planar charged system in a 
perpendicular 
magnetic field such that a new type of correlated 
ground state occurs. The Hall resistance develops 
plateaus at quantized values in the vicinity of the 
filling factor or statistics, $\nu$, which is related 
to the fraction of electrons that forms collective 
excitations as quasiholes or quasiparticles. Excitations 
above the Laughlin ground state are characterized, 
therefore, by $\nu$ and so we propose a 
{\it new hierarchy scheme} for the FQHE which gives us the 
possibility of {\it predicting} for which values 
of $\nu$ FQHE can be observed.
 
Our scheme is based on the intervals of definition 
of spin, $s$, for 
fractional spin particles which are related 
to the Hausdorff 
dimension, $h$. We verify that for some 
experimental values of $\nu$ for which the FQHE 
was observed the Hausdorff dimension is a rational 
number with an odd denominator 
( like the filling factor ). Thus, bearing in 
mind the condition, $1< h <2$, we can determine 
for which values of $h$, the statistics give 
numbers in the intervals of definition, as follows:  
       
\begin{eqnarray}
&&h_{1}=2-\nu,\;\;\;\; 0 < \nu < 1;\;\;\;\;\;
 h_{2}=\nu,\;\;\;\;\;
\;\;\;\;\; 1 <\nu < 2;\;\nonumber\\
&&h_{3}=4-\nu,\;\;\;\; 2 < \nu < 3;\;\;\;\;\;
h_{4}=\nu-2,\;\;\; 3 < \nu < 4;\;\nonumber\\
&&h_{5}=6-\nu,\;\;\;\; 4 < \nu < 5;\;\;\;\;
h_{6}=\nu-4,\;\;\;\; 5 < \nu < 6;\;\\
&&h_{7}=8-\nu,\;\;\;\; 6 < \nu < 7;\;\;\;\;\;
h_{8}=\nu-6,\;\;\; 7 < \nu < 8;\;\nonumber\\
&&h_{9}=10-\nu,\;\;8 < \nu < 9;\;\;\;
h_{10}=\nu-8,\;\; 9 < \nu < 10;\nonumber\\ 
&&etc.\nonumber
\end{eqnarray}

In these formulas, $h_{i}$, represents the Hausdorff 
dimension of 
the collective excitations which are characterized 
by the statistics $\nu$. 
This {\it hierarchy scheme} confirms our observation 
that{\it when the particles are interacting, in the 
presence of Chern-Simons field, 
the Hausdorff dimension changes because 
the statistics of the collective excitations 
change}\cite{R1}. Now, we give for some 
experimental values of $\nu$ the respective 
values of $h$, that is, 

\begin{eqnarray}
&&( h\;,\; 0< \nu < 1):\\
&&\left(\frac{9}{5},\frac{1}{5}\right),\;\left(\frac{12}{7},
\frac{2}{7}\right),\; 
\left(\frac{5}{3},\frac{1}{3}\right),\;
\left(\frac{8}{5},\frac{2}{5}\right),\nonumber\\
&& \left(\frac{11}{7},\frac{3}{7}\right),\;
\left(\frac{14}{9},\frac{4}{9}\right),\;
\left( \frac{13}{9},\frac{5}{9}\right),\;
\left( \frac{10}{7},\frac{4}{7}\right),\nonumber\\
&&\left(\frac{7}{5},\frac {3}{5}\right),\;
\left(\frac{4}{3},\frac{2}{3}\right),\; 
\left(\frac{6}{5},\frac{4}{5}\right);\nonumber\\ 
&&( h\;,\; 1 < \nu < 2 ):\\
&&\left(\frac{5}{3},\frac{5}{3}\right),\;
\left(\frac{13}{7},\frac{13}{7}\right),\;
\left(\frac{13}{9},\frac{13}{9}\right),\nonumber\\
&&\left(\frac{10}{7},\frac{10}{7}\right),\;
\left(\frac{7}{5},\frac{7}{5}\right),\;
\left(\frac{4}{3},\frac{4}{3}\right);\nonumber\\
&&( h\;,\; 2 < \nu < 3 ):\\
&&\left(\frac{4}{3},\frac{8}{3}\right),\;
\left(\frac{5}{3},\frac{7}{3}\right).\nonumber
\end{eqnarray}

\noindent We can see another 
interesting point from these pairs of numbers: Some 
collective excitations with different spins have the same 
value of $h$, that is, the nature of the occurrence 
of FQHE for that values of $\nu$ can be classified in terms 
of $h$, so we can say that this number 
classifies {\it the collective excitations in 
terms of its homotopy class}\cite{R3}. In this 
way, the Laughlin 
wavefunctions\cite{R4} can be understood as a mapping between 
homotopy classes of the collective excitations.

\acknowledgments
I would like to thank Steven F. Durrant for reading
 the manuscript.


\begin{thebibliography}{99}
\bibitem{R1} W. da Cruz, preprint/UEL-DF/W-01/97 
(submitted to Phy. Rev. D. ); W. da Cruz, 
preprint/UEL-DF/W-02/97; W. da Cruz, 
preprint/UEL-DF/W-03/97 ( hep-th/9802135 ).
\bibitem{R2} A. M. Chang in {\it The Quantum Hall Effect}, 
ed. by R. E. Prange and S. M. Girvin, ( Springer Verlag, 1987) 
and references therein. 
\bibitem{R3} A. Lerda, {\it Anyons}, Lectures 
Notes in Physics, ( Springer Verlag, 1992); S. Forte, 
Rev. Mod. Phys. {\bf 64}, 193 (1992) and references therein.
\bibitem{R4} R. B. Laughlin, Phys. Rev. Lett.{\bf 50} 
1395 (1983); ibid, Phys. Rev. {\bf B23} 3383 (1983); 
B. I. Halperin, Phys. Rev. Lett. {\bf 52} 1583 (1984); 
F. D. M. Haldane, Phys. Rev. Lett. 
{\bf 51} 605 (1983); R. Tao and D. J. Thouless, Phys. Rev. 
{\bf B28} 1142 ( 1983); P. W. Anderson, Phys. Rev.
{\bf B28} 2264 (1983); J. K. Jain, Adv. Phys. 
{\bf 41}, 105 (1992) 
and references therein; J. E. MacDonald and 
F. D. M. Haldane, Phys. Rev. {\bf B55}, 7818 
(1997); R. M\'elin and B. Dou\c{c}ot, Phys. Rev. 
{\bf B55} 13135 (1997) and references therein.   
\end{thebibliography}
\end{document}